\documentclass{article}

\usepackage{microtype}
\usepackage{graphicx}
\usepackage{subcaption}
\usepackage{ulem}
\usepackage{booktabs}
\usepackage{multirow}
\usepackage{xcolor}
\usepackage{pifont}

\usepackage{xurl}
\usepackage{hyperref}

\usepackage[accepted]{icml2026}

\usepackage{amsmath}
\usepackage{amssymb}
\usepackage{mathtools}
\usepackage{amsthm}
\usepackage{enumitem}
\usepackage{placeins}
\usepackage{afterpage}

\usepackage[capitalize,noabbrev]{cleveref}

\theoremstyle{plain}

\theoremstyle{definition}

\theoremstyle{remark}

\newcommand{\method}{RepetitionCurse}

\begin{document}

\twocolumn[
  \icmltitle{\method: Measuring and Understanding Router Imbalance in Mixture-of-Experts LLMs under DoS Stress}

  \icmlsetsymbol{equal}{*}

  \begin{icmlauthorlist}
    \icmlauthor{Ruixuan Huang}{hkust}
    \icmlauthor{Qingyue Wang}{hkust}
    \icmlauthor{Hantao Huang}{ntu}
    \icmlauthor{Yudong Gao}{hkust}
    \icmlauthor{Dong Chen}{hkust}
    \icmlauthor{Shuai Wang}{hkust}
    \icmlauthor{Wei Wang}{hkust}
  \end{icmlauthorlist}

  \icmlaffiliation{hkust}{HKUST, Hong Kong}
  \icmlaffiliation{ntu}{NTU, Singapore}

  \icmlcorrespondingauthor{Qingyue Wang}{qingyue.wang@ust.hk}

  \icmlkeywords{Mixture-of-Experts, Router Imbalance, Denial-of-Service Attack, Expert Parallelism}

  \vskip 0.3in
]

\printAffiliationsAndNotice{}

\begin{abstract}
    Mixture-of-Experts architectures have become the standard for efficient LLM scaling, typically employing expert parallelism to distribute experts across devices. However, the absence of explicit load balancing constraints during inference allows adversarial inputs to trigger severe routing concentration. We demonstrate that out-of-distribution prompts can manipulate the routing mechanism such that tokens are routed to a small, shared subset of top-$k$ experts, which creates computational bottlenecks on certain devices while forcing others to idle. This converts an efficiency mechanism into a denial-of-service attack vector, leading to violations of service-level agreements for time-to-first-token (TTFT). We propose \method, a black-box strategy to exploit this vulnerability. By identifying a universal flaw in MoE router behavior, \method~constructs attack prompts using simple repetitive token patterns in a model-agnostic manner. On widely deployed MoE models hosted on 8-GPU clusters, our method increases TTFT by 20\% to 148\%, significantly degrading service quality.
\end{abstract}

\section{Introduction}

While the capabilities of large language models (LLMs) continue to advance rapidly~\cite{DBLP:journals/tkde/CaiJWTKH25,10.1145/3735632}, deploying dense models at scale incurs prohibitive computational and memory costs. To address this challenge, the Mixture-of-Experts (MoE) architecture~\cite{DBLP:conf/iclr/ShazeerMMDLHD17} has emerged as a practical solution by employing dynamic routing mechanisms to select a subset of independent experts for each token during inference.
To deploy MoE models under fixed resource budgets, industry typically uses expert parallelism (EP), hosting distinct experts on separate devices~\cite{du2021glam0,fedus2021switch} to reduce inter-GPU communication and improve memory efficiency.

\begin{figure}[t]
    \centering
    \includegraphics[width=\linewidth]{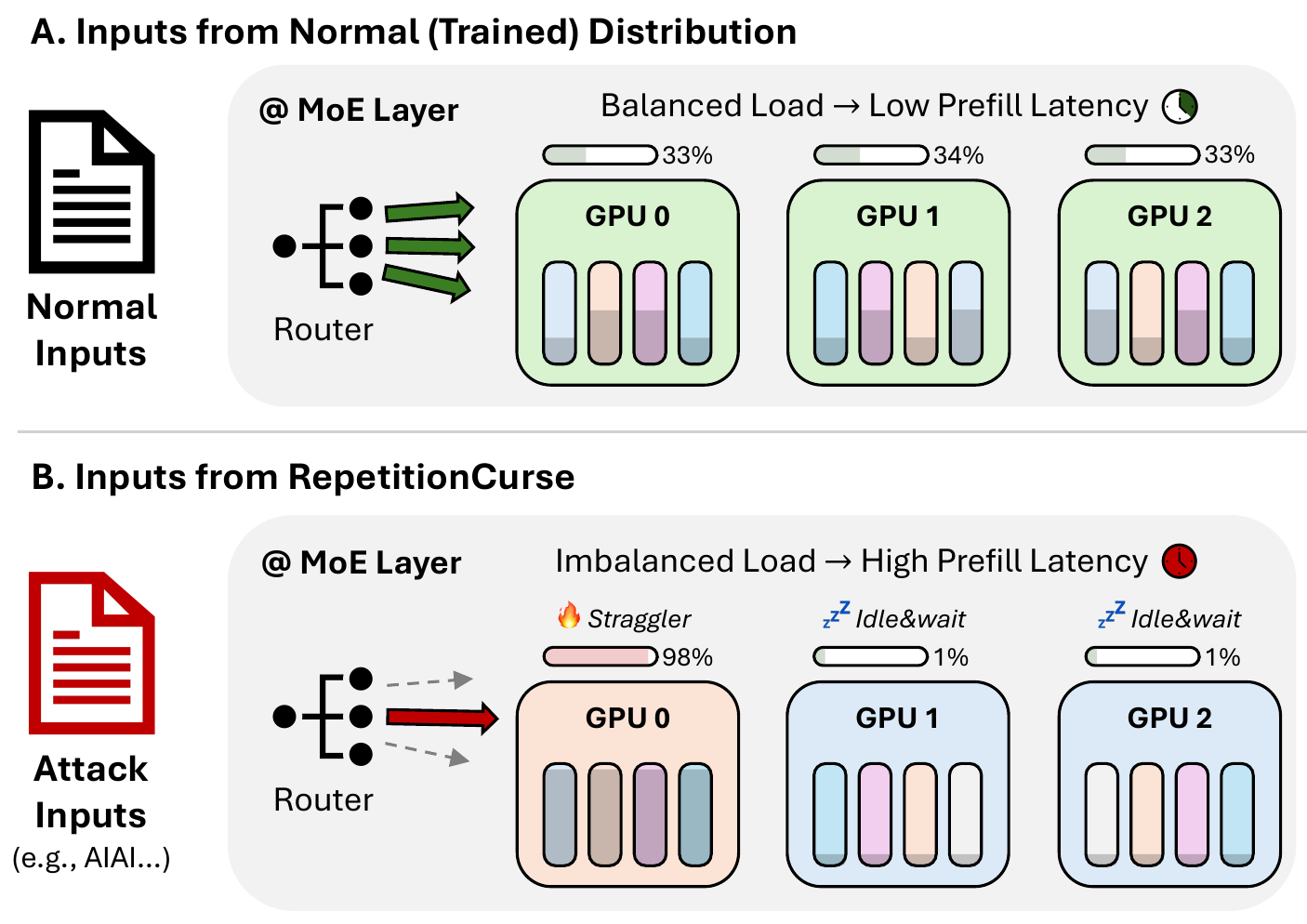}
    \caption{Impact of expert routing on GPU utilization. We use 3-way expert parallelism ($\text{EP size}=3$), where three GPUs are used with four experts deployed on each GPU. The shaded regions represent their workload.}
    \label{fig:overview_attack}
\end{figure}

However, without explicit balancing constraints during inference, EP deployment exposes a system-level vulnerability:  \textit{imbalanced routing might leave some GPUs idle, waiting for stragglers and increasing latency.}
As shown in Figure~\ref{fig:overview_attack}, benign workloads distribute load evenly across three GPUs~\cite{alephalpha2025deepseek}, but adversarial inputs (generated by our method) bias routing, making GPU 0 a bottleneck while idling the others and increasing latency.
We view such routing behavior as a deployment-level denial-of-service (DoS) attack: rather than forcing excessive generation of long outputs~\cite{li2025thinktrap0,zhang2024crabs0}, such an attack induces resource underutilization and delayed execution. In production settings, this latency amplification can systematically violate \textit{service-level agreements} (SLA), a contractual guarantee of reliable service delivery~\cite{gong2025past}, triggering financial penalties, unnecessary autoscaling, and service degradation for benign users.

In this paper, we propose \method, a simple yet highly effective method to employ prompts consisting of repetitive tokens to induce such a DoS attack. 
This is motivated by the discrepancy in balancing strategy between training and inference in MoE models. During training, MoE models use expert- and device-level balance losses~\cite{deepseek-ai2024deepseek0v20,deepseek-ai2024deepseekv3} to evenly distribute tokens across experts; while during inference, we uncover that out-of-distribution inputs can disrupt this balance, causing stragglers and idle devices. Leveraging this, \method~requires no access to model parameters nor any prior knowledge of routing mechanisms, and operates effectively in black-box settings. By forcing synchronized stalls, an attacker can degrade the \textit{time-to-first-token} (TTFT, \citet{DBLP:conf/osdi/AgrawalKPMKGTR24}), a key metric for interactive LLM serving, for all legitimate users within the same batch.

We investigate 139 MoE configurations with the most downloads on Huggingface and identify that \textbf{nearly all current MoE architectures are susceptible to this vulnerability}. 
Our experiments show that under the common 8-GPU EP deployment, TTFT latency can be increased by 1.29$\times$ to 2.48$\times$ on models such as Mixtral-8x7B and Qwen3-30B-A3B, breaking the $P_{99}$ SLA guarantee as the violation rate rises from 1\% to 1.4\% $\sim$ 13.6\% and turning the system's parallelism against itself. 
We uncover that higher degrees of EP substantially increase the system's susceptibility to such attacks (Section~\ref{sec:RQ1}). Accordingly,
we recommend that LLM service providers, pending the introduction of better inference-time load-balancing strategies, carefully limit the EP size when deploying MoE models.
Our contributions are summarized as follows:
\begin{itemize}[leftmargin=*]
    \item We identify a critical vulnerability in EP-based MoE serving systems. We are the first to demonstrate that routing imbalance can be weaponized to launch DoS attacks.
    \item We propose \method, a black-box attack method that uses repetitive tokens to amplify prefilling latency and degrade time-to-first-token of MoE serving systems.
    \item We conduct a comprehensive measurement across multiple MoE architectures, quantifying the severity of latency amplification and revealing key system-level factors.
\end{itemize}

\section{Related Work}

\noindent\textbf{MoE Models and Expert Parallel.} MoE architecture scales model capacity while maintaining low inference costs~\cite{DBLP:conf/iclr/ShazeerMMDLHD17}. Following the success of GPT-4~\cite{openai2023gpt04} and Mixtral~\cite{jiang2024mixtral}, MoE has become a standard for LLMs. To support these models, popular inference engines like vLLM~\cite{kwon2023efficient} and SGLang~\cite{zheng2024sglang} utilize Expert Parallelism (EP, \citet{du2021glam0,fedus2021switch}), which enables scaling of expert numbers across multiple devices with minimal communication overhead.

\noindent\textbf{DoS Attacks on LLM Systems.} LLMs deployed as inference systems are susceptible to DoS attack vectors. \citet{gao2024denial0of0service,zhang2024crabs0} explore how to make LLMs continuously generate until reaching the maximum limit of tokens to exhaust limited backend system resources. \citet{zhang2025leechhijack0,zhou2026beyond} study how to secretly occupy the computing resources of emerging AI agent systems through backdoors or prompt injections. \citet{li2025thinktrap0} investigate inducing models with thinking capabilities to engage in endless reasoning processes, leading to resource exhaustion. Unlike existing works where attackers must pay for every token produced to occupy the system, our method sabotages the actual compute-efficiency of each token.
\section{Background}

\subsection{Inference of Transformer-based LLMs}

The Transformer architecture serves as the backbone for current LLMs~\cite{vaswani2017attention,sun2025efficient}. The body of an LLM is constructed by stacking multiple Transformer blocks. Each block typically comprises a self-attention layer and a feed-forward network (FFN) layer. MoE models replace the dense FFN layers with MoE layers.

The inference of Transformer-based LLMs proceeds in two phases: (1) \textit{Prefill}: The model processes the entire input sequence batch in parallel, computing the initial KV cache. The prefill phase is {compute-bound} due to its heavy computational load of processing the full input sequence. (2) \textit{Decoding}: The model generates output tokens sequentially in an autoregressive manner. The decoding phase is \textit{memory-bound} due to the high memory bandwidth overhead incurred by frequent access to the KV cache.
Modern inference engines, such as vLLM and SGLang, employ prefill-decoding disaggregation strategies~\cite{DBLP:conf/osdi/ZhongLCHZL0024} to allocate these phases to distinct hardware resources for improved performance, making latency particularly sensitive to any inefficiency in the prefill computation.

\subsection{Service-Level Objective \& Agreement}

LLM serving relies on time-to-first-token (TTFT) as a critical metric of responsiveness~\cite{patke2024queue}, which measures the duration from the request arrival to the generation of the first output token.

Interactive applications typically impose tight \textit{service-level objectives} (SLOs) on TTFT to ensure the user experience. For example, a common target for production chatbots is the $P_{99}$ TTFT of less than 20s~\cite{gnewuch2022opposing}, meaning that 99\% of user requests must receive their first token within this timeframe. Service providers guarantee SLOs through service-level agreements (SLAs) through contractual obligations that stipulate financial penalties for performance failures. Therefore, DoS attacks that inflate TTFT will cause SLA violations, jeopardizing service reliability and incurring significant costs.

\subsection{MoE Models with Expert Parallelism}
\begin{figure}[t]
    \centering
    \includegraphics[width=\linewidth]{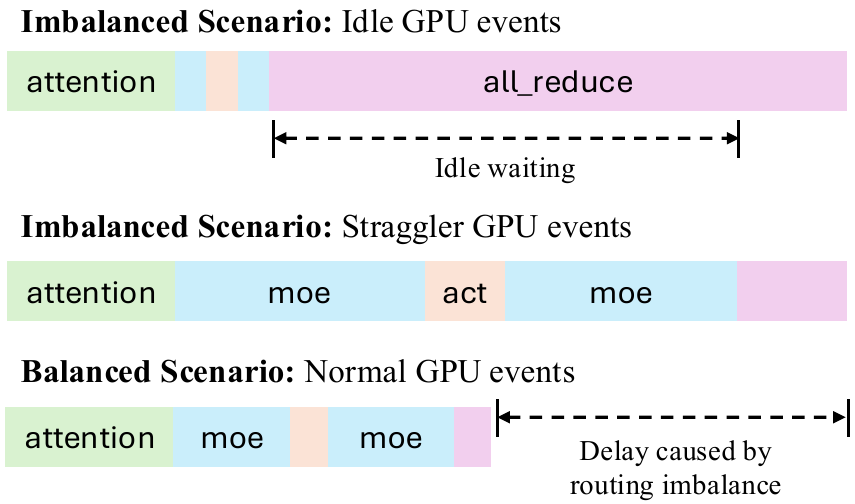}
    \caption{GPU event timeline from profiling under both balanced and imbalanced routing for a single MoE layer computation. Unrelated events such as normalization are omitted.}
    \label{fig:events}
\end{figure}

Distinct from dense models, an MoE layer comprises multiple weight-independent experts. For each input token, only a small subset of these experts is activated. Considering a model containing $E$ experts per layer and input hidden state $h$, one MoE layer first determines the participation of experts via the gating logits provided by the router:
\begin{equation}
    \label{eq:gating}
    G(h) = \text{Softmax}(h \cdot W_{\text{router}}) \in \mathbb{R}^E
\end{equation}
where $W_{\text{router}}$ is the weight of the router. Then, the top-$k$ experts (denoted as $\mathcal{K}$) with the highest gating logits are selected to process the token:
\begin{equation}
    y = \sum_{e_i \in \mathcal{K}} G(h)_{e_i} \cdot e_i(h)
\end{equation}
where $G(\cdot)_{e_i}$ is the logit score of the $i$-th expert. Under EP, the complete weights of one or more experts are hosted on a single GPU. Modern inference engines employ fused kernels, such as grouped GEMM~\cite{vllm_fused_moe_modular_kernel_docs}, to jointly execute the computation for all experts residing on a GPU. Thus, kernel execution time is dictated by the cumulative workload per GPU. Inference execution under EP is highly sensitive to routing distribution~\cite{he2025capacity0aware}. When router imbalance occurs, GPUs with lighter loads must idle and wait for the straggler GPU to complete before initiating the inter-GPU all-reduce synchronization. Figure~\ref{fig:events} illustrates the GPU event timeline under both balanced and imbalanced routing for a single MoE layer computation.

\section{Attack Methodologies}

\subsection{Threat Model}

\noindent \underline{\textbf{Asset Owner:}} In this paper, we define the asset owner as the service provider who deploys MoE-based LLMs on a multi-GPU cluster with EP and prefill-decoding disaggregation strategies to serve users via public APIs. For example, DeepSeek AI uses an EP size of 32 to deploy DeepSeek-V3 prefilling~\cite{deepseek-ai2024deepseekv3}. 

\noindent \underline{\textbf{Attack Objective:}} By submitting adversarial requests, the attacker aims to launch DoS attacks to inflate the TTFT for legitimate user requests, and to degrade the service's responsiveness and reliability, causing SLA violations. The attacker disregards the quality of the model's output, as the attack vectors target the computational costs during the prefill phase of inference.

\noindent \underline{\textbf{Attacker Capabilities:}} The attacker, operating in black-box scenarios~\cite{fu-etal-2024-vulnerabilities}, is permitted to send input queries to the API or the chatbot entry while possessing no knowledge of the backend architecture.

\subsection{Attack Prompt Formulation}
\label{sec:prompt_formulation}

To create a computational bottleneck, the attack prompt must manipulate the MoE router to consistently select the same top-$k$ experts across the entire sequence. Since routing is a deterministic function of token embeddings, such concentration can be achieved by minimizing the difference between consecutive embeddings.

We characterize this objective through the lens of the embedding space.
Let $h^l_i(X) \in \mathbb{R}^{d}$ denote the hidden state of the $i$-th token at layer $l$ given a prompt $X = [x_1, \dots, x_N]$, and let $\bar{h}^l(X) = \frac{1}{N}\sum_{i=1}^{N} h^l_i(X)$ denote the per-layer centroid.
The router's ability to balance load implicitly relies on the assumption that these hidden states are semantically diverse and well-distributed across the embedding space.
We quantify the dispersion of token representations entering the router via the per-layer embedding variance:
\begin{equation}
\label{eq:emb_var}
D(H^l(X)) = \frac{1}{N}\sum_{i=1}^{N} \big\| h^l_i(X) - \bar{h}^l(X) \big\|_{2}^{2}.
\end{equation}
Under this view, the attacker's optimal prompt corresponds to the solution of Equation (\ref{eq:attack_obj}), which is a prompt that collapses the layer-wise hidden representations toward a near-degenerate cluster, leaving the router unable to distribute load across experts.
\begin{equation}
\label{eq:attack_obj}
X^{*} = \arg\min_{X} \sum_{l=1}^{L} D(H^l(X)),
\end{equation}
Solving Equation~\eqref{eq:attack_obj} requires white-box access to layer-wise hidden states and gradients, which is unavailable under our black-box threat model. Therefore, we introduce \method, a streamlined, gradient-free attack that constructs prompts by \textbf{repeating identical tokens} (excluding the instruction template and system prompt) as an empirical approximation to Equation~\eqref{eq:attack_obj}: identical tokens drive consecutive hidden states toward a near-degenerate cluster, simultaneously minimizing embedding variance and the empirical embedding entropy~\cite{skeanlayer}. \method~induces routing collapse onto a fixed expert subset without requiring computationally expensive optimization~\cite{hayes2024buffer}. This approach is practically optimal and context-independent (Section~\ref{sec:length_scalability}), and can bypass KV caching by simply modifying the leading token.

\subsection{Attack Performance Upper Bound Analysis}
\label{sec:attack_performance}

We derive the theoretical limit of load imbalance caused by \method~under a specific deployment configuration. Consider an MoE model with a vocabulary $\mathcal{V}$. Let $\mathcal{E}_l = \{e_{l,1}, \dots, e_{l,E}\}$ denote the set of experts at the $l$-th layer. The model is deployed across a set of devices denoted as $\mathcal{D} = \{d_1, \dots, d_M\}$. Based on EP strategy, we define the Expert-GPU mapping $\mathcal{M}_l: \mathcal{D} \to 2^{\mathcal{E}_l}$, representing a deployment configuration. 
Assuming each GPU hosts $E_d=|\mathcal{M}_l(d)|$ experts and the routing strategy is top-$k$, the theoretical maximum imbalance (TMI), defined as the ratio of the worst-case single-device load $\min(k,E_d)$ to the ideal balanced load $k/|\mathcal{D}|$:
\begin{equation}
\label{eq:tmi}
\text{TMI} = \frac{\min(k, E_d)}{k/|\mathcal{D}|}
\end{equation}
Intuitively, the TMI represents the multiplicative factor by which the straggler's workload exceeds its fair share. 
Equation (\ref{eq:tmi}) reveals two distinct regimes. For sparser models like DeepSeek-V3 ($k \le E_d$), the attacker can achieve the perfect bottleneck when scaling the EP size. For models like Mixtral-8x7B under high-degree parallelism (e.g., $|\mathcal{D}|=\text{EP size}=8 \Rightarrow E_d=1<k=2$), the activated experts are physically forced to span multiple devices. Hence, the bottleneck is capped at $|\mathcal{D}|\cdot E_d/k=4$, instead of $8$.

\subsection{Constraints on Practical Attack Performance}

We identify two factors that constrain the attack performance, preventing it from reaching TMI.

\noindent \textbf{Obliviousness to Expert-GPU Mapping Change.}
TMI represents a static worst-case attack assumption where all target experts are hosted within a minimal number of GPUs. In practice, dynamic mechanisms like Expert Parallelism Load Balancer (EPLB, \citet{deepseek_eplb}) periodically change Expert-GPU mapping $\mathcal{M}$ based on observed workloads. While such relocation can mitigate imbalance, it is performed at coarse time scales (e.g., every 10 mins) and incurs non-negligible overhead; within each relocation window, the mapping remains static. Therefore, in this paper, we adopt the default mapping where experts are assigned to GPUs in order, which is default for major inference engines such as vLLM and SGLang when EPLB is not enabled. 

\begin{figure}[t]
    \centering
    \includegraphics[width=\linewidth]{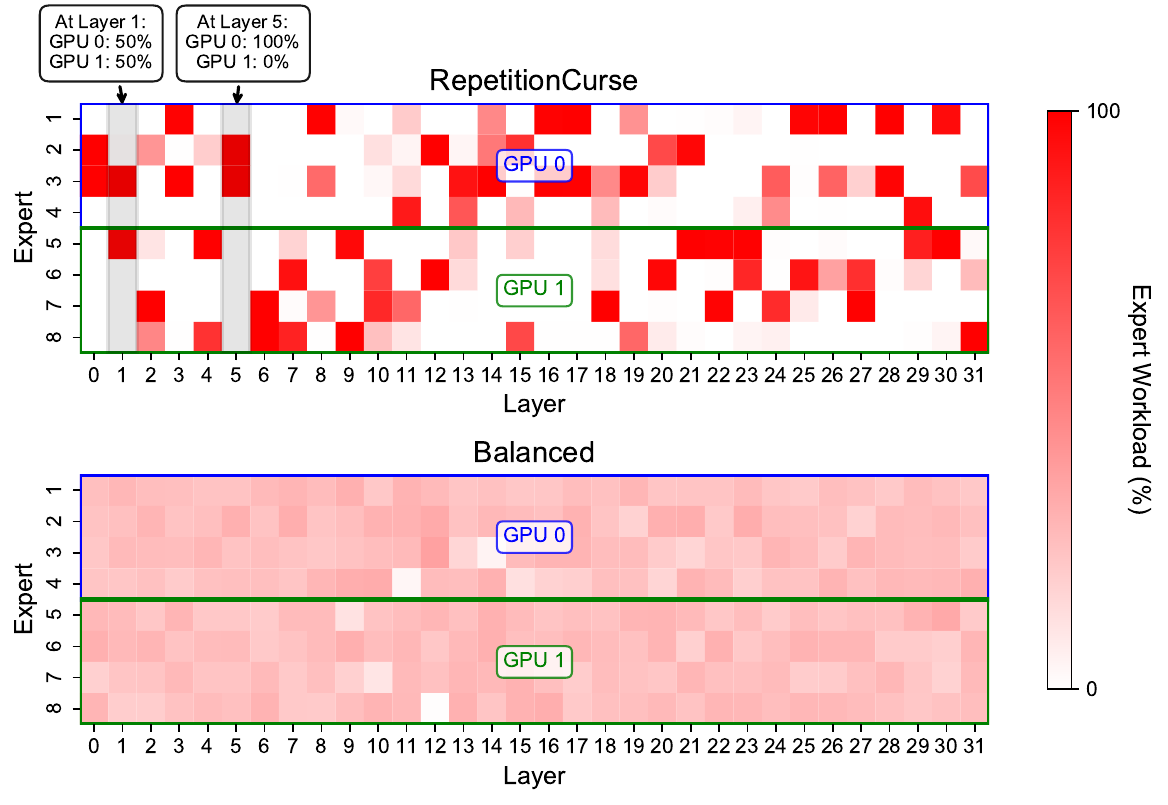}
    \caption{Expert workload comparison between \method~and balanced baseline. Input prompts for balanced baseline are sampled from LongBench~\cite{bai2024longbench2}. Each cell represents the percentage of tokens routed to the expert.}
    \label{fig:optimality_heatmap}
\end{figure}

\textbf{Inability to Control Target Experts.}
While \method~can achieve black-box routing of all tokens to the same top-$k$ experts, it provides no leverage to manipulate the target experts. Figure~\ref{fig:optimality_heatmap} presents a comparative visualization of the routing distribution between \method~samples and a balanced baseline on Mixtral-8x7B ($E=8,k=2$). It demonstrates that the router maintains a relatively uniform distribution on natural text, while \method~induces extreme workload concentration. Yet, under $\text{EP size}=2$, this concentration doesn't necessarily manifest as a tangible latency bottleneck. For layers where the two attacked experts are placed on different GPUs (e.g., Layer 1), the workload remains evenly split across devices, and no effective delay is incurred despite concentrated routing. However, under the unknown Expert-GPU mapping assumption, this limitation effectively degenerates into a probabilistic factor. The attack's performance depends on the stochastic alignment between the target experts and their physical location.

\begin{table}[t]
\centering
\caption{Model architectures and attack-related configurations.}
\label{tab:coverage}
\resizebox{0.9\linewidth}{!}{
\begin{tabular}{@{}l cccc@{}}
\toprule
\textbf{Model} 
& $L$ 
& $E$ 
& Top-$k$ 
& $|\mathcal{V}|$ \\
\midrule

Mixtral-8x7B          & 32 & 8   & 2  & 32000 \\
Mixtral-8x7B-It & 32 & 8   & 2  & 32000 \\
Mixtral-8x7B-Chinese  & 32 & 8   & 2  & 57000 \\
Mixtral-8x7B-Nous     & 32 & 8   & 2  & 32002 \\
\midrule

Qwen3-30B-A3B                & 48 & 128 & 8  & 151936 \\
Qwen3-30B-A3B-It       & 48 & 128 & 8  & 151936 \\
Qwen3-Coder-30B-A3B-It & 48 & 128 & 8  & 151936 \\
\midrule

GPT-OSS-20B  & 24 & 32  & 4 & 201088 \\
GPT-OSS-120B & 36 & 128 & 4 & 201088 \\
\midrule

Llama-4-Scout-17B-16E-It & 48 & 16  & 1 & 202048 \\
DeepSeek-V2-Lite               & 27 & 64  & 6 & 102400 \\
Kimi-Linear-It $^*$          & 27 & 256 & 8 & 163840 \\
Kimi-Linear-Base $^*$          & 27 & 256 & 8 & 163840 \\

\bottomrule
\end{tabular}
}
\begin{flushleft}
\small
$*$ The model has linear attention mechanism.
\end{flushleft}
\end{table}

\section{Experiments}

In this section, we conduct a comprehensive empirical measurement of the proposed \method~attack. Our experiments are organized from three perspectives of the attack:

\noindent \textbf{RQ1:} Is the routing imbalance vulnerability universal across distinct model vocabularies and MoE architectures?

\noindent \textbf{RQ2:} How does the attack impact inference latency and SLA compliance in production environments?

\noindent \textbf{RQ3:} Can optimized Expert-GPU mapping effectively mitigate this vulnerability under black-box scenarios? 

Through these evaluations, we aim to uncover the fundamental trade-offs between MoE parallelization efficiency and system robustness. We introduce a set of novel metrics specifically designed to quantify adversarial workload imbalance for real-world impact measurement.

\subsection{MoE Models Investigation \& Setup}

To measure the universality of this vulnerability across different models, we scrape 139 configuration files of MoE models from Huggingface with over 1,000 downloads, analyzing their expert configurations and routing strategies. We summarize the percentage of different MoE architectures and their average downloads in Table~\ref{tab:moe_models} in Appendix~\ref{sec:moe_models}. The results show that over 50\% of the surveyed models are Mixtral-like models, which are characterized by a smaller $E$ (i.e., $\le 8$). The remaining models with architectures like DeepSeekV3 and Qwen3Moe are characterized by sparser activated experts, with typically only 2 to 12 experts activated, out of $E$ ranging from 128 to 512.

Motivated by the investigation, we focus on 13 popular MoE models in the experiments, including 4 Mixtral-like low-sparse models and other 9 high-sparse models. Table~\ref{tab:coverage} summarizes their expert configurations and routing strategies. The selected models collectively cover both base and post-trained variants, different attention mechanisms (standard and linear), and a wide range of expert cardinalities from 8 to 256, enabling a systematic analysis of attack robustness across training stages and architectural choices.

\subsection{RQ1: Vulnerability Universality Investigation}
\label{sec:RQ1}

For this RQ, we investigate the universality of the vulnerability introduced by \method~across two distinct dimensions: (1) \textit{Cross-vocabulary}, assessing whether the entire vocabulary can be exploited to construct attack prompts; and (2) \textit{Cross-model}, examining whether this vulnerability persists across diverse MoE architectures.

\noindent \textbf{Metrics.} To quantify how broadly the vocabulary $\mathcal{V}$ can be exploited to construct the attack prompts, we introduce a coverage metric to characterize the vulnerability across the entire $\mathcal{V}$ under a specific mapping $\mathcal{M}$.
For a prompt $P_t$ constructed by \method~using token $t \in \mathcal{V}$, let $\rho_e(P_t)$ represent the percentage of tokens routed to expert $e$. We calculate the computational load $L_{l,d}(P_t)$ for device $d$ at layer $l$, and define the bottleneck score $B(P_t)$ as:
\begin{equation}
\left\{
    \begin{aligned}
        L_{l,d}(P_t) &= \frac{1}{|\mathcal{M}_l(d)|} \sum_{e \in \mathcal{M}_l(d)} \rho_e(P_t) \\
        B(P_t) &= \frac{1}{L}\sum_{l=1}^{L} \max_{d \in \mathcal{D}} L_{l,d}(P_t)
    \end{aligned}
\right.
\end{equation}
where $B(P_t) \in [0,1]$ measures the degree of routing imbalance induced by prompt $P_t$, and a larger value indicates a stronger concentration of token $t$ on a single device. Furthermore, we define the average bottleneck  
\(
\mathcal{B} = \frac{1}{|\mathcal{V}|} \sum_{t \in \mathcal{V}} B(P_t),
\)
which directly reflects the probability that a randomly sampled token from $\mathcal{V}$ 
can induce near-maximal routing concentration.

\begin{figure}[t]
    \centering
    \includegraphics[width=\linewidth]{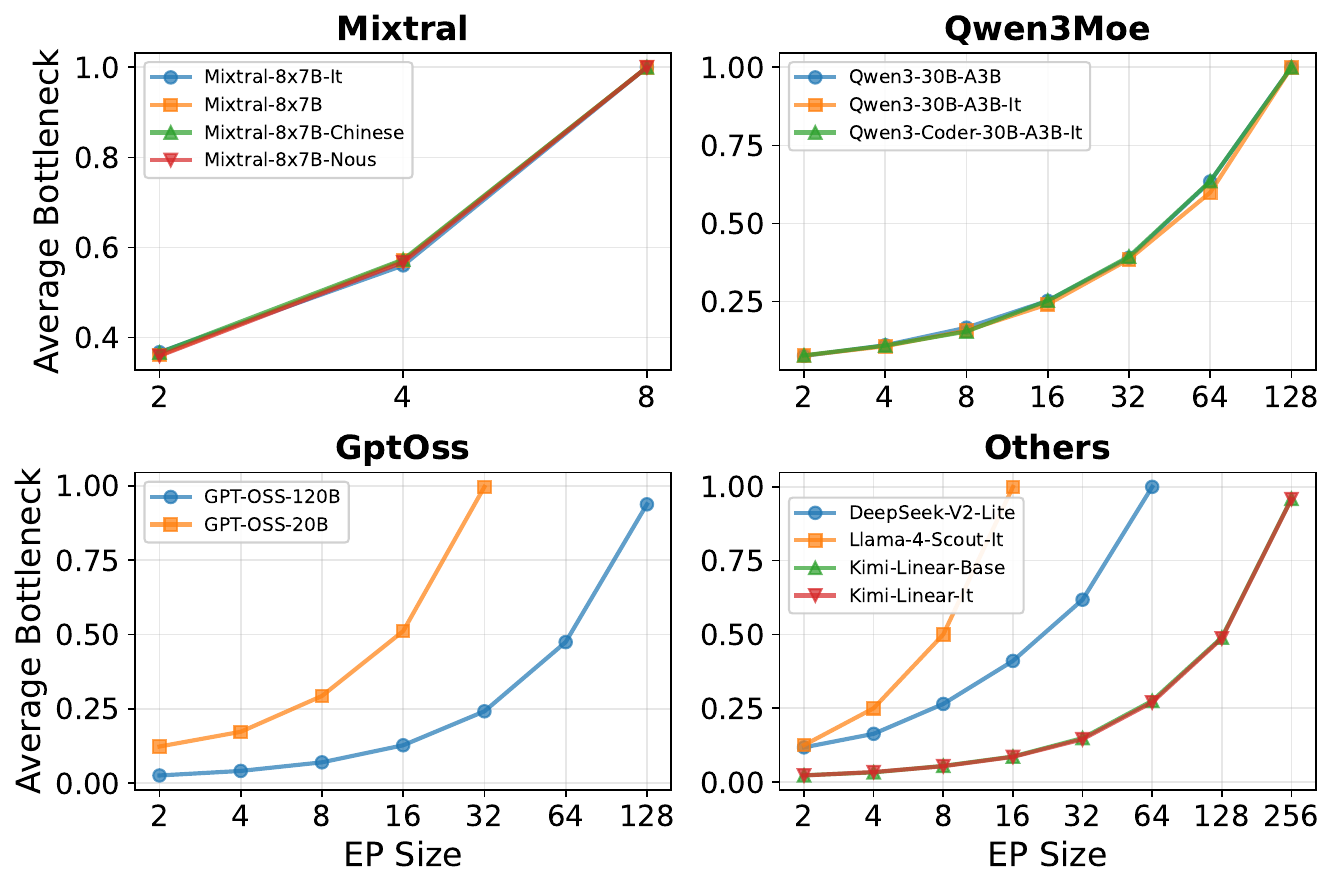}
    \caption{Coverage of different models under different EP size.}
    \label{fig:bottleneck_coverage}
\end{figure}

\noindent \textbf{Results.} Figure~\ref{fig:bottleneck_coverage} presents the average bottleneck coverage $\mathcal{B}$ of different models under different EP size. When $\text{EP size} = E$, bottleneck coverage approaches 1 for all models. This implies that nearly any token in the vocabulary can be used to construct an effective attack prompt. It demonstrates the broad universality of the vulnerability across both vocabularies and models.

The average bottleneck coverage consistently increases with the EP size, indicating that larger EP size exacerbates the vulnerability during the prefill phase. This creates a deployment dilemma: \textbf{increasing EP size to improve efficiency simultaneously amplifies the attack surface}. Under commonly used EP settings (e.g., 8–32), models with a larger $E$ (e.g., Qwen3-30B-A3B series) exhibit lower bottleneck coverage. Under the same parameter budget, a shallow and wide MoE model (large $E$, small $L$) is more resistant to our attack compared to a deep and narrow one. We also find that models from the same family exhibit similar coverage trends (e.g., Mixtral-8x7B and its fine-tuned variants). This suggests that the vulnerability is primarily introduced during pretraining rather than post-training or instruction tuning.

\noindent \textbf{Answer to RQ1:} The vulnerability introduced by \method~is intrinsic and universal. We find that nearly any token in the vocabulary can serve as a trigger for routing concentration. This is a shared architectural feature across diverse MoE models regardless of attention mechanisms and training stage, advocating a trade-off between parallelization efficiency and system robustness.

\begin{table*}[htbp]
\centering
\caption{The LAR ($\text{LAR}_{\text{moe}}$ and $\text{LAR}_{\text{ttft}}$) of different models under different EP size. All results are reported as $x/y$ for $\alpha=\frac12 / 1$.}
\label{tab:latency_moe}
\resizebox{\linewidth}{!}{%
\begin{tabular}{@{}l ccccc ccccc@{}}
\toprule
\textbf{Model} 
  & \multicolumn{5}{c}{$\text{LAR}_{\text{moe}}$ under EP size =} 
  & \multicolumn{5}{c}{$\text{LAR}_{\text{ttft}}$ under EP size =} \\
\cmidrule(lr){2-6} \cmidrule(lr){7-11}
 & 2 & 4 & 8 & 16 & 32 & 2 & 4 & 8 & 16 & 32 \\
\midrule

Mixtral-8x7B          & 1.38 / 1.52 & 1.54 / 1.93 & 2.01 / 2.68 & $\circ$ & $\circ$ & 1.07 / 1.23 & 1.31 / 1.75 & 1.61 / 2.48 & $\circ$ & $\circ$ \\
Mixtral-8x7B-It       & 1.38 / 1.67 & 1.39 / 2.02 & 1.94 / 3.12 & $\circ$ & $\circ$ & 1.09 / 1.32 & 1.22 / 1.74 & 1.65 / 2.48 & $\circ$ & $\circ$ \\
Mixtral-8x7B-Chinese  & 1.34 / 1.52 & 1.48 / 2.04 & 2.12 / 2.73 & $\circ$ & $\circ$ & 1.03 / 1.16 & 1.30 / 1.64 & 1.44 / 2.14 & $\circ$ & $\circ$ \\
Mixtral-8x7B-Nous     & 1.29 / 1.72 & 1.51 / 2.32 & 1.81 / 2.85 & $\circ$ & $\circ$ & 1.06 / 1.27 & 1.34 / 1.81 & 1.57 / 2.29 & $\circ$ & $\circ$ \\

\midrule

Qwen3-30B-A3B          & 1.17 / 1.45 & 1.24 / 1.73 & 1.42 / 1.89 & 1.76 / 2.59 & 2.28 / 3.22 & 1.07 / 1.14 & 1.10 / 1.19 & 1.12 / 1.22 & 1.32 / 1.76 & 1.53 / 2.15 \\
Qwen3-30B-A3B-It       & 1.18 / 1.51 & 1.27 / 1.78 & 1.43 / 1.91 & 1.81 / 2.77 & 2.16 / 3.14 & 1.07 / 1.15 & 1.08 / 1.21 & 1.12 / 1.22 & 1.27 / 1.78 & 1.49 / 2.01  \\
Qwen3-Coder-30B-A3B-It & 1.14 / 1.52 & 1.26 / 1.66 & 1.33 / 1.69 & 1.82 / 2.53 & 2.32 / 3.04 & 1.06 / 1.15 & 1.08 / 1.20 & 1.13 / 1.20 & 1.29 / 1.72 & 1.51 / 2.08  \\

\midrule

GPT-OSS-20B & 1.04 / 1.14 & 1.12 / 1.34 & 1.20 / 1.46 & $\times$ & $\times$ & 1.03 / 1.07 & 1.06 / 1.17 & 1.08 / 1.20 & $\times$ & $\times$ \\
GPT-OSS-120B & 1.02 / 1.07 & 1.12 / 1.34 & 1.18 / 1.35 & $\times$ & $\times$ & 1.01 / 1.04 & 1.07 / 1.17 & 1.07 / 1.22 & $\times$ & $\times$ \\

\midrule

Kimi-Linear-It           & 1.02 / 1.12 & 1.06 / 1.28 & 1.21 / 1.51 & 1.44 / 1.98 & 1.65 / 2.41 & 1.01 / 1.08 & 1.04 / 1.12 & 1.10 / 1.28 & 1.18 / 1.37 & 1.32 / 1.63 \\
Kimi-Linear-Base           & 1.04 / 1.13 & 1.07 / 1.30 & 1.28 / 1.48 & 1.40 / 2.01 & 1.63 / 2.28 & 1.01 / 1.09 & 1.05 / 1.16 & 1.12 / 1.29 & 1.21 / 1.44 & 1.31 / 1.67 \\
DeepSeek-V2-Lite          & 1.13 / 1.37 & 1.29 / 1.59 & 1.48 / 1.95 & 2.03 / 3.10 & $\times$ & 1.04 / 1.09 & 1.08 / 1.14 & 1.10 / 1.20 & 1.45 / 1.97 & $\times$ \\
Llama-4-Scout-17B-16E-It & 1.13 / 1.25 & 1.41 / 1.89 & 1.49 / 2.11 & $\times$ & $\times$ & 1.03 / 1.08 & 1.24 / 1.48 & 1.50 / 1.91 & $\times$ & $\times$ \\

\bottomrule
\end{tabular}%
}
\begin{flushleft}
\small
$\circ$ Not applicable, because $E < \text{EP size}$; $\times$ Not applicable, because vLLM doesn't support this EP size for this model.
\end{flushleft}
\end{table*}

\subsection{RQ2: Real-world Latency Impact}
\label{sec:latency_metrics}
\begin{figure}[t]
    \centering
    \includegraphics[width=\linewidth]{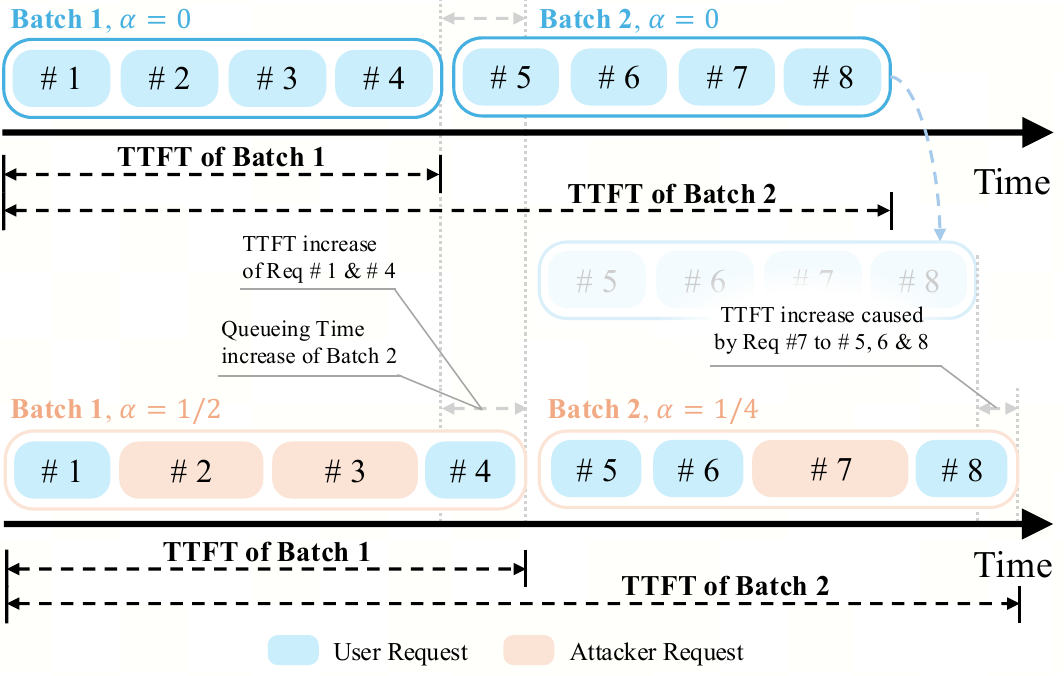}
    \caption{Impact of attacker requests on user TTFT and on subsequent batch queueing.}
    \label{fig:alpha_RQ2}
\end{figure}

For this RQ, we measure the real-world latency impact of the \method~attack. We focus on the latency amplification under two representative DoS attack scenarios: \ding{182} \textbf{Mixed-user batches.} Benign users submit requests concurrently with attackers and are served within the same batch. As a result, they experience the same inflated TTFT as the attacker. The slow completion of these batches further propagates to later requests, introducing additional queuing delays for subsequent users. \ding{183} \textbf{Attack-only batches.} Only attackers submit requests during certain periods. Although no benign users are directly affected within these batches, the attack degrades the system’s prefill capacity, causing subsequent users to wait longer and suffer increased TTFT. We provide an illustration in Figure~\ref{fig:alpha_RQ2} for how attack prompts affect the TTFT and queueing time of user requests.

\noindent \textbf{Setup.} In practice, we use $\alpha$ to parameterize the attack intensity, defined as the fraction of attack samples in each batch. Under this formulation, Setting~\ding{182} corresponds to $0<\alpha < 1$, while Setting~\ding{183} corresponds to $\alpha = 1$. We deploy the target models locally using vLLM with profiling. Following existing works~\cite{wu2025tokenlake0,xie2025strata0}, we collect 2048 user requests from ShareGPT~\cite{sharegpt2023} with token length between 1,000 and 5,000, and replace $\alpha\in\{0,\frac12,1\}$ proportion of them with attack prompts. Each batch contains 16,384 tokens. For each target model, we evaluate under $\text{EP size} = 2, 4 \text{ and } 8$. For several supported models, we also evaluate under $\text{EP size} = 16 \text{ and } 32$. For $\text{EP size} \le 8$, we utilize a single server with 8 NVIDIA A800 GPUs. For $\text{EP size} = 16$ and $32$, we extend the setup to 2 and 4 such nodes, respectively.

\noindent \textbf{Metrics.} We examine the impact of \method~attack on inference \textit{latency amplification ratio} (LAR) of both the TTFT and the MoE kernel execution time.  Let $T^i_{\text{moe}}$ denote the MoE kernel execution time of layer $i$ on the straggler GPU, we report $\text{LAR}_\text{moe}$ and $\text{LAR}_\text{ttft}$ as: 
\begin{equation}
    \label{eq:lar_moe}
    \left.
    \begin{aligned}
    \text{LAR}_\text{moe}^\alpha &= \frac{\sum_i T^i_{\text{moe}}(P_{\alpha})}{\sum_iT^i_{\text{moe}}(P_{\alpha=0})} \\
    \text{LAR}_\text{ttft}^\alpha &= \frac{\text{TTFT}(P_{\alpha})}{\text{TTFT}(P_{\alpha=0})}
    \end{aligned}
    \right.
\end{equation}
where $\text{LAR}_\text{moe}$ provides a direct measurement of the latency amplification within the MoE layers due to router concentration; while $\text{LAR}_\text{ttft}$ evaluates the systemic consequence, representing the increase in end-to-end TTFT.

\noindent \textbf{Results.} The results are shown in Table~\ref{tab:latency_moe}. We report results of $\alpha=\frac12\text{ and }1$ to demonstrate the batch latency and the queueing latency. Aligning with the trend of bottleneck coverage, the LARs increase with the increase of EP size. For Mixtral models, the deployment with $\text{EP size}=8$ increases TTFT by up to $148\%$; for sparser models like Qwen3-30B-A3B, the deployment with $\text{EP size}=32$ increases TTFT by up to $115\%$. Besides, we find that the \method~attack can significantly degrade the SLA guarantee. For example, under the common setting of 8-GPUs, our attack degrades the SLA guarantee from $P_{99}$ to between $P_{98.6}$ to $P_{86.4}$, which means serious damage to the service quality (detailed in Appendix~\ref{app:sla-vio}).

\noindent \textbf{Answer to RQ2:} \method~induces severe degradation in system responsiveness, turning the system's parallelism against itself. We observe a direct correlation between deployment scale and attack severity: higher EP sizes and lower expert sparsity $k$ significantly amplify latency, causing TTFT delays that systematically violate SLAs.

\subsection{RQ3: Expert-GPU Mapping Effect}

For this RQ, we focus on the performance of the proposed attack when the backend Expert-GPU mapping is unknown. We investigate this from a defensive perspective: if a defender can proactively identify vulnerable experts that are susceptible to acting as routing attractors, they can devise an optimal mapping strategy to minimize computational hotspots. By simulating the vulnerability-aware mapping, we can investigate if the vulnerability is an intrinsic architectural flaw that persists even under the most resilient deployment.

\noindent \textbf{Setup \& Metrics.} We perform a vocabulary-wide scan for each expert across the target models. Specifically, we count the number of unique tokens in the vocabulary that, when used to construct a \method~prompt, result in the expert receiving over a threshold $=90\%$ of the prompt's tokens. This metric identifies vulnerable experts that act as attractors for a disproportionate amount of repetitive tokens. Let $v_{l,e}$ denote the number of trigger tokens corresponding to expert $e$ in layer $l$. We compute
\begin{equation}
    \label{eq:exp_entropy}
    N_{\text{eff}}=\exp(H_l) = \exp\left(-\sum_{e=1}^{E} p_{l,e} \log p_{l,e}\right)
\end{equation}
where $p_{l,e} = v_{l,e} / |V|$, to quantify the effective number of experts $N_{\text{eff}}$ serving as attractors across the vocabulary.

\begin{figure}[t]
    \centering
    \begin{subfigure}{\linewidth}
        \centering
        \includegraphics[width=\linewidth]{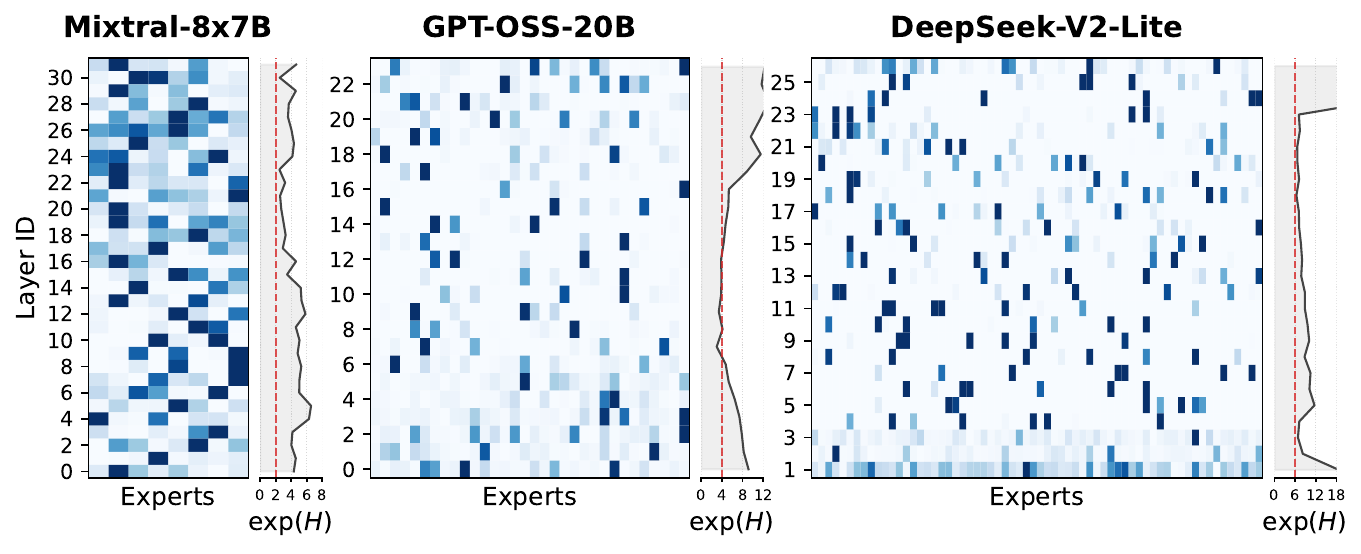}
    \end{subfigure}
    \begin{subfigure}{\linewidth}
        \centering
        \includegraphics[width=\linewidth]{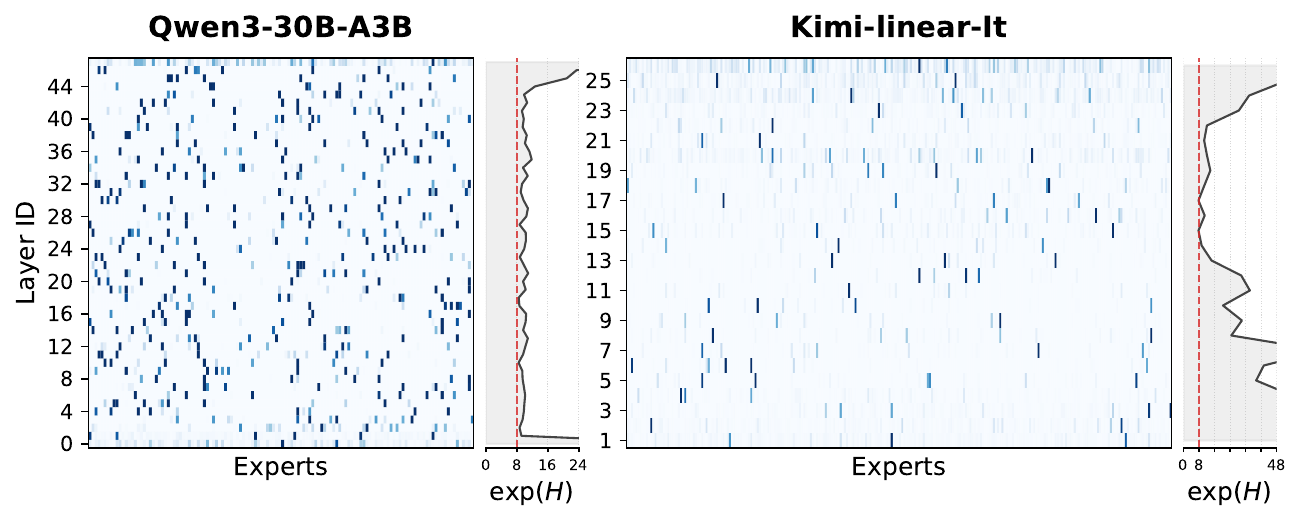}
    \end{subfigure}
    \begin{subfigure}{\linewidth}
        \centering
        \includegraphics[width=0.9\linewidth]{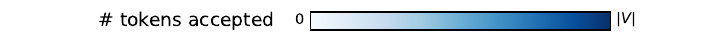}
    \end{subfigure}
    \caption{Vulnerable experts distribution. Each cell represents the number of trigger tokens of this expert. We provide $\exp(H)$ per layer to illustrate the number of effective attraction experts.}
    \label{fig:vulnerable_experts_distribution}
\end{figure}

Figure \ref{fig:vulnerable_experts_distribution} presents the resulting heatmaps for five selected models. We observe a compelling and consistent pattern across these models: experts in the early and late layers generally exhibit broad vulnerability, where trigger tokens are distributed relatively evenly. In contrast, the intermediate layers show significant sparsity, where vulnerability is confined in only a sparse subset of experts. This trend is most pronounced in Qwen3-30B-A3B, where the effective number of experts in intermediate layers consistently approaches $k$. Conversely, it exhibits a sudden increase in the first three and final layers, indicating a transition from high concentration to a more dispersed distribution.

Consequently, the most resilient Expert-GPU mapping to this attack is to deploy these vulnerable experts separately on different GPUs. Under such a placement, even if an attack prompt successfully activates all vulnerable experts simultaneously, the resulting computation is distributed across the cluster, preventing the bottleneck and mitigating LAR. Under the assumption of an unknown mapping, this strategy can measure the lower bound of the \method~attack.

\begin{figure}[t]
    \centering
    \includegraphics[width=\linewidth]{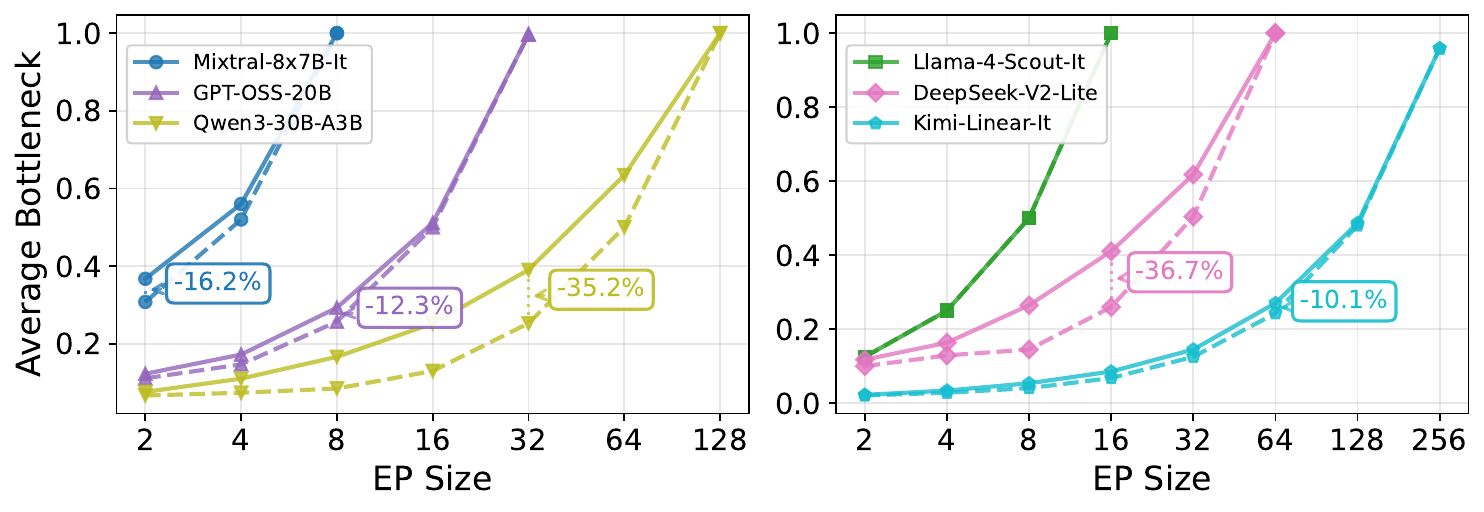}
    \caption{Effect of vulnerability-aware load balance strategy. Solid lines show the bottleneck coverage under the default mapping, while dashed lines show coverage applying the proposed strategy.}
    \label{fig:all_models_coverage_comparison}
\end{figure}

\noindent \textbf{Results.} To evaluate the effectiveness of this strategy, we simulate EP deployments using a greedy load-balancing algorithm (See Algorithm~\ref{alg:load_balance} in Appendix~\ref{app:RQ3-alg}) that assigns experts to GPUs with their number of trigger tokens. Figure~\ref{fig:all_models_coverage_comparison} shows the resulting average bottleneck coverage $\mathcal{B}$ after rebalancing and the differences. We observe that this defense is particularly effective for configurations with a moderate $E$ (e.g., 64–128) and large enough $k$, where vulnerable experts can be separated across devices. For example, for DeepSeek-V2-Lite ($E=64,k=6$), its coverage decreased by 36.7\% at $\text{EP size}=16$. However, for larger EP sizes and for models with small $k$ (e.g., Llama-4-Scout with $k=1$), any mapping strategy is ineffective.

\textbf{Answer to RQ3:} Expert-GPU mapping offers only conditional mitigation. While identifying and distributing attractor experts can reduce bottlenecks in models with moderate EP size, this defense collapses for highly sparse models or high EP size. In these cases, the vulnerability persists regardless of the mapping strategy, confirming that the mitigation effect imposed by unknown Expert-GPU mapping in black-box scenarios is limited.

\section{Extended Analysis and Implications}

\subsection{Length \& Context Robustness}

\label{sec:length_scalability}
A natural robustness issue for the vulnerability exploited by \method~is its sensitivity to prompt length and context: whether inducing routing concentration requires long attack prompts, and whether similar effects arise under different contexts?

\noindent \textbf{Setup.} We investigate this on two MoE models: Mixtral-8x7B and Qwen3-30B-A3B, spanning prompt lengths ranging from 100 to 16,384 tokens and five different system prompts whose lengths range from 10 to 5,000. To characterize routing concentration behavior, we measure the normalized entropy ${H}(P)$ of the expert selection distribution averaged across all layers. This metric captures the routing behavior without relying on specific deployments. Formally, the normalized entropy is defined as:
\begin{equation}
\label{eq:entropy}
{H}(P) = \frac{1}{L} \sum_{l=1}^{L} \frac{-\sum_{e \in \mathcal{E}_l} \rho_{l,e}(P) \log \rho_{l,e}(P)}{\log E}
\end{equation}
where $\rho_{l,e}(P)$ represent the percentage of tokens routed to expert $e$ of layer $l$. Lower entropy means more concentrated expert selection, indicating stronger routing imbalance.

\begin{figure}[t]
    \centering
    \includegraphics[width=\linewidth]{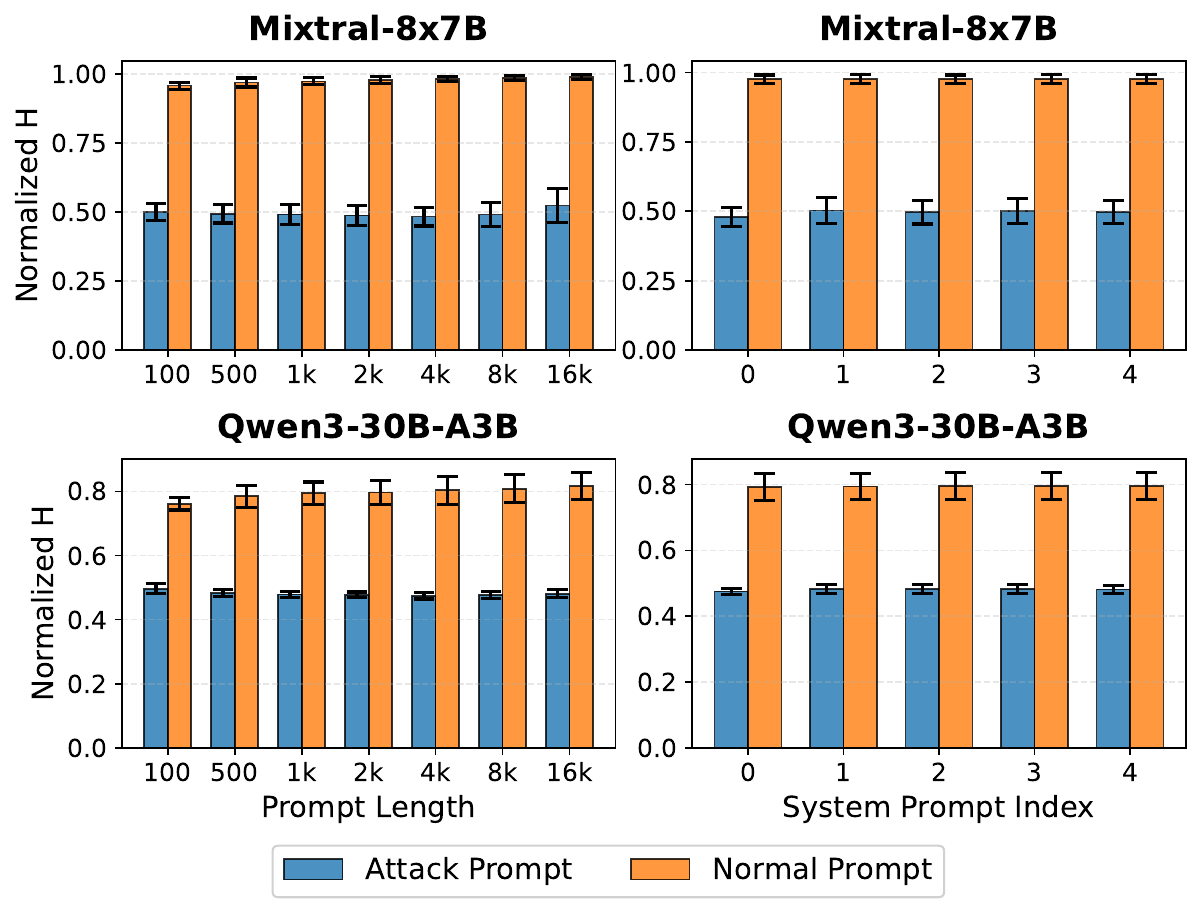}
    \caption{Average normalized entropy of the expert selection distribution under different prompt lengths and system prompts.}
    \label{fig:length_scalability}
\end{figure}

\noindent \textbf{Results.} The results shown in Figure~\ref{fig:length_scalability} demonstrate that \method~exhibits length and context robustness. We observe that the induced token concentration remains consistent across the entire range of evaluated prompt lengths. 

Furthermore, this behavior proves to be invariant to the diversity of system prompts. Once an attacker identifies the optimal adversarial tokens, they can consistently use those tokens to construct prompts, requiring only minor adjustments to the system prompt, without the attack's effectiveness being compromised by KV caching. This consistency also indicates that the vulnerability revealed by \method~is systemic to the nature of MoE architectures. When tokens exhibit repetitive patterns, their corresponding embeddings are immediately dominated by this repetition, despite the layer-wise attention and position encoding. Consequently, the router is manipulated into deterministically assigning these tokens to a fixed set of experts.

\subsection{End-to-End API LAR}

From the perspective of a black-box attacker, we evaluate the \method~attack on commercial LLM APIs. In this scenario, the DoS impact can only be measured within its specific request batch. To ensure no severe disruption to these services, we conduct the attack at a low request rate with minute-wise intervals.

\noindent \textbf{Setup \& Metrics.} We measure the TTFT for API requests to indicate the actual performance degradation caused by \method.  We define the LAR of the API as:
\begin{equation}
    \text{LAR}_\text{api} = \frac{\text{TTFT}(P_{\text{attack}})}{\text{TTFT}(P_{\text{normal}})}
\end{equation}
For each API, we construct 100 attack prompts and 100 normal prompts, each with a length of 20,000. We set \texttt{max\_new\_tokens} to 1 to get the TTFT.

\begin{table}[t]
\centering
\caption{Comparison of average TTFT between \method~and normal prompts, sorted in descending order by $\text{LAR}_\text{api}$.}
\label{tab:time_perf_comparison}
\resizebox{\linewidth}{!}{
\begin{tabular}{@{}l ccc@{}}
\toprule
\multirow{2}{*}{\textbf{Model}} & \multicolumn{2}{c}{{$T(P)$ (s)}} & $\text{LAR}_\text{api}$   \\
\cmidrule(lr){2-3}

 & {attack} & {normal} & \small{(95\% CILB)} \\
\midrule

Kimi-Linear-Instruct & 14.595 & 2.066 & 4.728 \\
DeepSeek-R1-Turbo & 10.762 & 1.942 & 4.012 \\
Qwen3-Coder-30B-A3B-Instruct & 17.067 & 3.454 & 3.233 \\
Mixtral-8x7B & 25.789 & 5.288 & 3.063 \\
GPT-OSS-20B & 6.197 & 1.672 & 2.399 \\
GPT-OSS-120B & 13.220 & 2.951 & 2.341 \\
Qwen3-30B-A3B & 8.080 & 2.569 & 2.138 \\
Llama-4-Scout-17B-16E & 3.866 & 2.002 & 1.514 \\

\bottomrule
\end{tabular}
}
\end{table}
\noindent \textbf{Results.} The results are shown in Table~\ref{tab:time_perf_comparison}. Considering potential network fluctuations, we calculate the point estimate of $\text{LAR}_\text{api}$ and report the 95\% confidence interval lower bound (CILB) to indicate latency amplification. The results demonstrate that inference requests constructed using \method~consistently incur higher average API TTFT than normal prompts, effectively resulting in a DoS effect. Across the majority of evaluated APIs, the observed LAR exceeds 100\% significantly, ranging from $1.514\times$ to $4.728\times$. This indicates that \method~enables substantially more efficient participation in DoS attacks against commercial LLM inference services.

\section{Conclusion}

In this paper, we uncover a routing imbalance vulnerability in MoE-based LLM serving systems that can be exploited for DoS attack. We demonstrate that our proposed RepetitionCurse can manipulate routers to create computational stragglers. As industry increasingly adopts highly sparse MoE models for efficiency, our findings underscore an urgent trade-off between parallelization and system robustness, calling for inference-time load-balancing strategies.

\section*{Impact Statement}

This paper highlights a critical system-level vulnerability in MoE LLM deployments that can be exploited to cause severe denial-of-service effects.
To mitigate the potential for misuse, we have adopted a responsible disclosure. The proposed RepetitionCurse is a black-box attack designed primarily as a diagnostic tool for researchers and system providers to evaluate the robustness of their serving infrastructure. As MoE architectures become the backbone of industrial LLM services due to their efficiency, our findings serve as a timely warning for cloud service providers.

\section*{Acknowledgments}

We sincerely thank the editors and the anonymous reviewers for their valuable feedback and guidance. This paper was supported in part by grants from the Research Grants Council of the Hong Kong Special Administrative Region, China (No. C6015-23G) and an ITF grant under the contract TS/161/24FP. We thank HKUST Fok Ying Tung Research Institute and National Supercomputing Center in Guangzhou Nansha Sub-center for computational resources.

\bibliography{custom}
\bibliographystyle{icml2026}

\newpage
\appendix
\onecolumn
\section*{Appendix}
\section{MoE Models Investigation}
\label{sec:moe_models}

To measure the universality of \method~across various MoE models, we conduct a configuration investigation of MoE models on Huggingface. We include 139 valid MoE models that have over 1,000 downloads and are non-quantized versions. The architectures of these models, the percentage of models corresponding to each architecture, and the average download count of models under each architecture category are presented in Table~\ref{tab:moe_models}. Notably, 54.68\% of the models adopt the Mixtral architecture, and other mainstream architectures of current MoE models include Qwen3Moe, DeepSeekV3 and DeepSeekV2, etc.

\begin{table}[htbp]
\centering
\caption{Summary of investigated MoE models on Huggingface.}
\label{tab:moe_models}
\begin{tabular}{@{}c l c | c l c@{}}
\toprule
\textbf{Percent} & \textbf{Architecture} & \textbf{Avg. Downloads}
& \textbf{Percent} & \textbf{Architecture} & \textbf{Avg. Downloads} \\
\midrule
54.68\% & Mixtral        & 9,834
& 1.44\% & Arctic        & 4,988 \\

13.67\% & DeepseekV3    & 139,431
& 1.44\% & Deepseek     & 9,336 \\

5.76\%  & Qwen3Moe      & 325,997
& 1.44\% & Qwen3Next    & 885,050 \\

5.76\%  & DeepseekV2    & 88,858
& 1.44\% & FlexOlmo     & 21,874 \\

2.88\%  & Olmoe         & 23,958
& 1.44\% & Qwen2Moe     & 27,285 \\

1.44\%  & Lfm2Moe       & 8,346
& 1.44\% & KimiLinear   & 157,719 \\

1.44\%  & Glm4Moe       & 4,832
& Others & \multicolumn{2}{c@{}}{Llama4, MiniMax, GPT-OSS, \dots} \\
\bottomrule
\end{tabular}
\end{table}

\section{SLA Violations Analysis}
\label{app:sla-vio}

To rigorously quantify the impact of latency overhead on SLAs, we model the TTFT distribution. Let $T$ be the random variable representing the TTFT. We assume $T$ follows a Log-Normal distribution, $T \sim \text{LogNormal}(\mu, \sigma^2)$, which is widely observed in service latency characteristics~\cite{wang2024burstgpt0}.

\begin{figure}[h]
    \centering
    \includegraphics[width=0.45\linewidth]{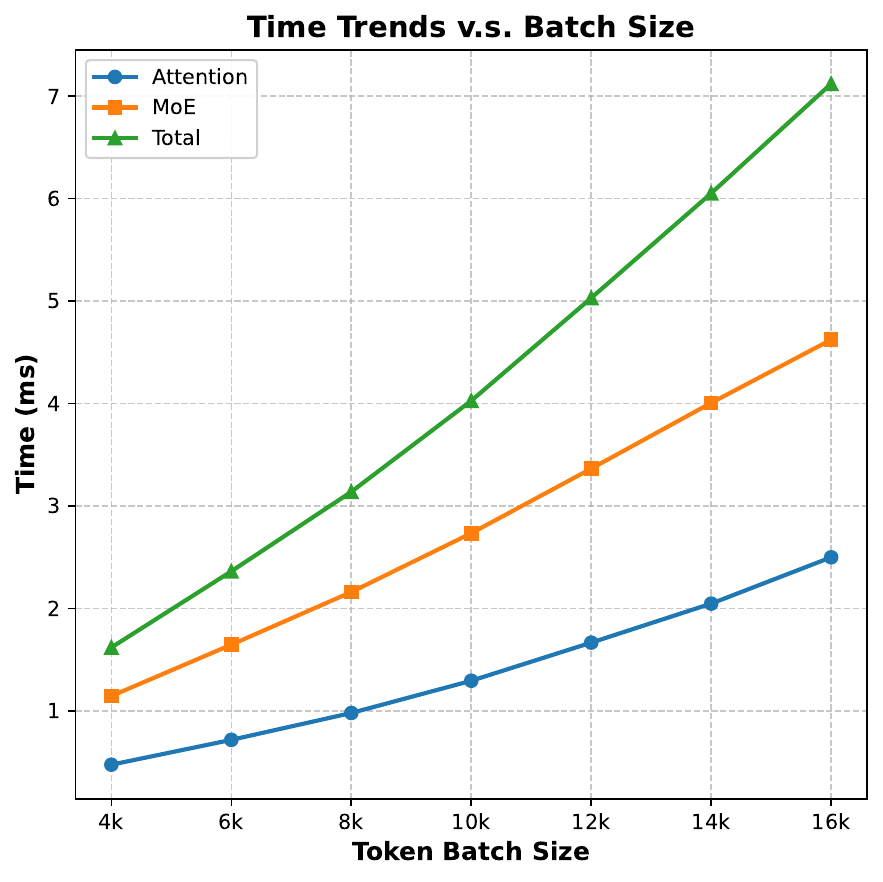}
    \caption{Breakdown of computation time of a layer for Qwen3-30B-A3B across different token batch sizes.}
    \label{fig:time_trend}
\end{figure}

A critical property of the Log-Normal distribution is that its shape parameter $\sigma$ is invariant under linear scaling. 
Although the self-attention mechanism theoretically exhibits complexity of $O(N^2)$, our profiling on Qwen3-30B-A3B (shown in Figure \ref{fig:time_trend}) reveals that within the operational context (batch size $\le \text{16,384}$), the total latency is dominated by the MoE computations with complexity of $O(N)$, where $N$ is the input prompt length.  
Consequently, the shape parameter of the TTFT distribution closely mirrors that of the input token distribution.

\begin{figure}[t]
    \centering
    \includegraphics[width=\linewidth]{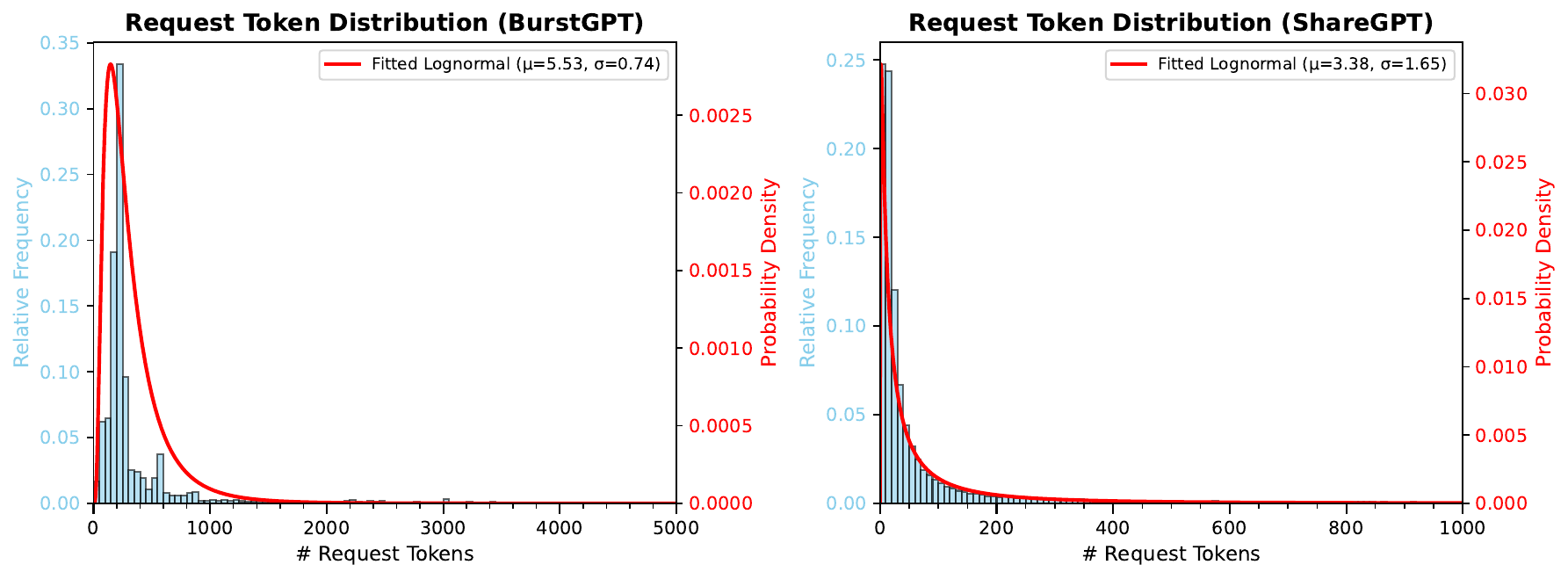}
    \caption{Empirical cumulative distribution of request tokens for BurstGPT (left) and ShareGPT (right) with fitted Log-Normal curves. The tight fit confirms that input distributions can be effectively modeled as Log-Normal, justifying the use of derived $\sigma$ values for SLA analysis.}
    \label{fig:sigma}
\end{figure}

We validate the distribution using two distinct datasets: BurstGPT~\cite{wang2024burstgpt0} and ShareGPT~\cite{sharegpt2023} representing real-world input prompt length distributions. As shown in Figure \ref{fig:sigma}, we fit Log-Normal distributions to the request token counts of both datasets. The empirical data shows a strong fit, yielding two distinct shape parameters. For BurstGPT, its $\sigma \approx 0.74$, indicating a more concentrated distribution; while for ShareGPT, its $\sigma \approx 1.65$, indicating a heavy-tailed distribution with high variance.

Now we analyze how the LAR induced by the attack will cause the original $P_{99}$ SLA to drop given the distribution of TTFT. The probability density function of $T$ is given by:
\begin{equation}
f(t) = \frac{1}{t\sigma\sqrt{2\pi}} \exp\left( -\frac{(\ln t - \mu)^2}{2\sigma^2} \right), \quad t > 0
\end{equation}

Let $\tau_{99}$ be the SLA threshold such that the original compliance probability is $P_{99} = 99\%$. This is defined by:
\begin{equation}
P_{99} = P(T \le \tau_{99}) = \Phi\left( \frac{\ln \tau_{99} - \mu}{\sigma} \right) = 0.99
\end{equation}
where $\Phi(\cdot)$ is the cumulative distribution function of the standard normal distribution. Let $Z_{99} = \Phi^{-1}(0.99) \approx 2.326$. It follows that $\mu = \ln \tau_{99} - \sigma Z_{99}$.

Consider a scenario where the system introduces an LAR of $k$ (e.g., $k=1.25$ implies a 25\% increase). The new latency variable is $T' = kT$. The new compliance probability $P'$ under the \textit{original} threshold $\tau_{99}$ is:
\begin{equation}
\label{eq:sla-deg}
P' = P(T' \le \tau_{99}) = P(kT \le \tau_{99}) = \Phi\left( Z_{99} - \frac{\ln k}{\sigma} \right)
\end{equation}

Equation (\ref{eq:sla-deg}) demonstrates that the degraded SLA probability $P'$ depends solely on the original safety margin $Z_{99}$, LAR $k$ and the TTFT distribution parameter $\sigma$.

Table \ref{tab:sla_degradation} presents the projected SLA degradation for observed LARs across different models, compared under the two workload distributions. The results highlight that workloads with lower variance are more sensitive to \method~attack. Under the BurstGPT distribution ($\sigma=0.74$), a $2.48\times$ latency increase causes the SLA compliance to decrease from 99\% to 86.4\%. In contrast, the heavy-tailed ShareGPT distribution ($\sigma=1.65$) is more robust, maintaining a 96.2\% compliance under the same degradation.

\begin{table}[htbp]
    \centering
    \caption{Projected SLA compliance probability degradation under different LARs on two real-world trace distributions: BurstGPT ($\sigma=0.74$) and ShareGPT ($\sigma=1.65$).}
    \label{tab:sla_degradation}
    \vspace{0.2cm}
    \begin{tabular}{lc|cc}
        \toprule
        & & \multicolumn{2}{c}{\textbf{Degraded SLA $ P'$}}\\
        \textbf{Model Example} & \textbf{LAR $k$} & {BurstGPT} & {ShareGPT} \\
        \midrule
        Mixtral-8x7B & 2.48 & $P_{86.4}$ & $P_{96.2}$ \\
        Llama-4-Scout & 1.91 & $P_{92.7}$ & $P_{97.3}$ \\
        Qwen3-30B-A3B & 1.22 & $P_{98.0}$ & $P_{98.6}$ \\
        \bottomrule
    \end{tabular}
\end{table}

\section{Vulnerability-Aware Load Balance Algorithm for RQ3}
\label{app:RQ3-alg}

Algorithm \ref{alg:load_balance} demonstrates how to utilize the results of vocabulary scan, i.e., the number of trigger tokens $v_{l,e}$ for each expert, to generate static defensive Expert-GPU mapping.

\begin{algorithm}[h]
\caption{Vulnerability-Aware Load Balance Strategy}
\label{alg:load_balance}
\begin{algorithmic}[1]

\REQUIRE 
Set of experts $\mathcal{E}_l = \{1, \dots, E\}$ at layer $l$, \\
Trigger token counts $\{v_{l,e}\}_{e \in \mathcal{E}_l}$, \\
Set of devices $\mathcal{D}$

\ENSURE 
Expert-GPU mapping $\mathcal{M}_l: \mathcal{D} \rightarrow 2^{\mathcal{E}_l}$

\STATE Initialize device load $L_d \leftarrow 0$ and expert count $N_d \leftarrow 0$ for all $d \in \mathcal{D}$
\STATE Calculate expert capacity per device $C \leftarrow E / |\mathcal{D}|$

\STATE Sort experts $e \in \mathcal{E}_l$ in descending order of trigger token counts $v_{l,e}$

\FOR{each expert $e$ in sorted order}
    \STATE Select $d^* = \arg\min_{d \in \mathcal{D}} L_d$ such that $N_d < C$
    \STATE Assign expert $e$ to device $d^*$: $\mathcal{M}_l(d^*) \leftarrow \mathcal{M}_l(d^*) \cup \{e\}$
    \STATE Update load and count: $L_{d^*} \leftarrow L_{d^*} + v_{l,e}$, $N_{d^*} \leftarrow N_{d^*} + 1$
\ENDFOR

\STATE \textbf{return} $\mathcal{M}_l$

\end{algorithmic}
\end{algorithm}

\section{Defense Proposals}
\label{app:defenses}

This section surveys the most natural defenses a practitioner would consider against \method, and discusses their applicability and limitations.

\noindent \textbf{Vulnerability-Aware Expert.} The structural defense already analyzed in Section~\ref{sec:RQ1} and Algorithm~\ref{alg:load_balance} places vulnerable experts on distinct devices to break the bottleneck. As shown in Figure~\ref{fig:all_models_coverage_comparison}, this defense is effective for moderate $E$ and large enough $k$, but collapses for highly sparse models or large EP sizes. We thus view it as a partial mitigation rather than a general solution.

\noindent \textbf{Perplexity-based Prompt Filtering.} A distinguishing feature of \method~prompts is their highly repetitive pattern, which yields very low perplexity (PPL) under any auxiliary language model. We measure the PPL of three representative prompt types using Llama-3-8B as the evaluator: (i) natural text from ShareGPT~\cite{sharegpt2023}, (ii) structured but non-repetitive text (code patches drawn from SWE-bench~\cite{jimenez2024swe0bench0}), and (iii) \method~prompts. Results are reported in Table~\ref{tab:defense_ppl}.

\begin{table}[htbp]
    \centering
    \caption{Perplexity of three prompt types across lengths, evaluated with Llama-3-8B.}
    \label{tab:defense_ppl}
    \vspace{0.2cm}
    \begin{tabular}{c ccc}
    \toprule
    \textbf{Length} & \textbf{Natural Text} & \textbf{Code} & \textbf{\method} \\
    \midrule
    50    & 76.4 & 11.8 & 1.98 \\
    512   & 15.8 &  6.3 & 1.10 \\
    2048  & 10.4 &  4.0 & 1.08 \\
    8192  &  9.4 &  3.8 & 1.21 \\
    \bottomrule
    \end{tabular}
\end{table}

While \method~prompts indeed exhibit the lowest PPL, deploying a PPL-threshold filter is impractical for two reasons. First, code-style prompts can also yield very low PPL. A threshold tight enough to flag \method~would also reject legitimate code-completion or structured-output traffic. Second, a PPL filter requires hosting an additional auxiliary evaluator on the request path, adding non-trivial GPU memory, network hops, and latency to every inference request, which directly contradicts the SLA goals the defense is meant to protect.

\section{Attack under Dynamic EPLB}
\label{app:eplb}

EPLB periodically rebalances the Expert-GPU mapping based on observed routing statistics. A natural question is whether such dynamic rebalancing is sufficient to neutralize \method. This section reports experiments comparing the bottleneck score $\mathcal{B}$ achieved under EPLB-derived mappings against our static, vulnerability-aware mapping (Algorithm~\ref{alg:load_balance}).

For each model and EP configuration, we evaluate two EPLB scenarios that capture how a real provider would observe the attack:

\noindent \textbf{Round 1 (R1, attack on benign-only EPLB).} We feed only benign traffic sampled from ShareGPT~\cite{sharegpt2023} to EPLB to derive the initial Expert-GPU mapping $\mathcal{M}^{(1)}$. We then launch \method~against $\mathcal{M}^{(1)}$ and report $\mathcal{B}$.

\noindent \textbf{Round 2 (R2, attack on adapted EPLB).} We feed a second batch of benign traffic that is now \emph{mixed} with attack prompts, letting EPLB re-derive a mapping $\mathcal{M}^{(2)}$ that has observed the attack pattern. We then re-launch \method~against $\mathcal{M}^{(2)}$ and report $\mathcal{B}$ again.

Table~\ref{tab:eplb} reports $\mathcal{B}$ for four representative models across multiple EP sizes. As shown, R1 isolates experts frequently activated by benign text. However, our attack targets vulnerable experts that may not be heavily loaded during normal traffic. Even when the system adapts by including attack traffic (R2), the score drops marginally. Alg.~\ref{alg:load_balance} outperforms EPLB. For models like Llama-4-Scout, the attack forces tokens onto a single expert, leaving the system completely defenseless. These results suggest that EPLB is insufficient to mitigate routing-level imbalance.

\begin{table}[htbp]
    \centering
    \caption{Bottleneck score $\mathcal{B}$ of \method~under dynamic EPLB mappings (R1 / R2) compared to Alg.~\ref{alg:load_balance}.}
    \label{tab:eplb}
    \vspace{0.2cm}
    \begin{tabular}{l c ccc}
    \toprule
    \textbf{Model} & \textbf{EP Size} & \textbf{EPLB R1} & \textbf{EPLB R2} & \textbf{Alg.~1} \\
    \midrule
    \multirow{2}{*}{Mixtral-8x7B}  & 2  & 0.36 & 0.33 & \textbf{0.30} \\
                                   & 4  & 0.56 & 0.54 & \textbf{0.52} \\
    \midrule
    \multirow{4}{*}{Qwen3-30B-A3B} & 8  & 0.15 & 0.14 & \textbf{0.08} \\
                                   & 16 & 0.25 & 0.22 & \textbf{0.13} \\
                                   & 32 & 0.38 & 0.33 & \textbf{0.25} \\
                                   & 64 & 0.59 & 0.52 & \textbf{0.50} \\
    \midrule
    \multirow{3}{*}{GPT-OSS-20B}   & 4  & 0.24 & 0.22 & \textbf{0.14} \\
                                   & 8  & 0.37 & 0.31 & \textbf{0.25} \\
                                   & 16 & 0.55 & 0.52 & \textbf{0.50} \\
    \midrule
    Llama-4-Scout                  & 4  & 0.25 & 0.25 & 0.25 \\
    \bottomrule
    \end{tabular}
\end{table}

\end{document}